\documentclass[aps,prl,twocolumn,superscriptaddress,showpacs,floatfix]{revtex4-2}

\usepackage{amsmath,amssymb,amsfonts,float,graphics,epsfig,epstopdf,color,verbatim,tabularx,bm,multirow,appendix}
\usepackage[utf8]{inputenc}
\usepackage[T1]{fontenc}
\usepackage{xcolor}
\usepackage{dsfont}
\usepackage{textcomp}
\usepackage{yfonts}
\usepackage{footnote}
\usepackage{bm}
\usepackage{subfigure}
\usepackage{mathrsfs}
\usepackage{graphicx}
\usepackage{verbatim}
\usepackage{hyperref}
\usepackage{multirow}
\usepackage{braket}
\usepackage[normalem]{ulem}
\usepackage{tikz}
\usetikzlibrary{calc}
\usetikzlibrary{shapes.multipart}

\renewcommand{\vec}[1]{{\mathbf #1}}

\newcommand{\comments}[1]{}

\newcommand{\stkout}[1]{\ifmmode\text{\sout{\ensuremath{#1}}}\else\sout{#1}\fi}

\makeatletter
\def\l@subsubsection#1#2{}
\makeatother

\begin{document}
	
	\title{Quantum criticality and entanglement for two dimensional long-range Heisenberg bilayer}
	
	\author{Menghan Song}
	\affiliation{Department of Physics and HKU-UCAS Joint Institute of Theoretical and Computational Physics, The University of Hong Kong, Pokfulam Road, Hong Kong SAR, China}
	
	\author{Jiarui Zhao}
	\email{jrzhao@connect.hku.hk}
	\affiliation{Department of Physics and HKU-UCAS Joint Institute of Theoretical and Computational Physics, The University of Hong Kong, Pokfulam Road, Hong Kong SAR, China}
	\author{Yang Qi}
	\affiliation{State Key Laboratory of Surface Physics, Fudan University, Shanghai 200438, China}
	\affiliation{Center for Field Theory and Particle Physics, Department of Physics, Fudan University, Shanghai 200433, China}
	\affiliation{Collaborative Innovation Center of Advanced Microstructures, Nanjing 210093, China}
	
	\author{Junchen Rong}
	\email{junchenrong@gmail.com}	
	\affiliation{Institut des Hautes \'Etudes Scientifiques, 91440 Bures-sur-Yvette, France}
	
	\author{Zi Yang Meng}
	\email{zymeng@hku.hk}
	\affiliation{Department of Physics and HKU-UCAS Joint Institute of Theoretical and Computational Physics, The University of Hong Kong, Pokfulam Road, Hong Kong SAR, China}
	
	\begin{abstract}

		The study of quantum criticality and entanglement in systems with long-range (LR) interactions is still in its early stages, with many open questions remaining to be explored. In this work, we investigate critical exponents and scaling of entanglement entropy (EE) in the LR bilayer Heisenberg model using large-scale quantum Monte Carlo (QMC) simulations. %and the recently developed nonequilibrium incremental algorithm for measuring EE. 
		By applying modified (standard) finite-size scaling above (below) the upper critical dimension and field theory analysis, we obtain precise critical exponents in three regimes: the LR Gaussian regime with a Gaussian fixed point, the short-range (SR) regime with Wilson-Fisher (WF) exponents, and a LR non-Gaussian regime where the critical exponents vary continuously from LR Gaussian to SR values. We compute the R\'enyi EE both along the critical line and in the N\'eel phase, and observe that as the LR interaction is enhanced, the area-law contribution in EE gradually vanishes both at quantum critical points (QCPs) and in the N\'eel phase. The log-correction in EE arising from sharp corners at the QCPs also decays to zero as LR interaction grows, whereas that for N\'eel states, caused by the interplay of Goldstone modes and restoration of the symmetry in a finite system, is enhanced. %as LR interaction becomes stronger. 
		Relevant experimental settings to detect these nontrivial properties for quantum many-body systems with LR interactions are discussed.
	\end{abstract}
	
	\date{\today}
	\maketitle
	
	\noindent{\textcolor{blue}{\it Introduction.}---}
	In recent years, research on long-range (LR) interacting quantum many-body systems has attracted significant attention %and intrigued numerous investigations 
	from the perspectives of statistical physics, renormalization group field theory and lattice model computation. These systems exhibit many exotic properties that awaits to be explored thoroughly, such as the modification of dynamic spectra~\cite{DiesselGeneralized2022,songDynamical2023,EddySpin2004,DefenuLong2021,Irenee2017}, violation of area law scaling of entanglement entropy (EE)~\cite{KoffelEntanglement2012,Zehan2021} and the breaking of Lieb-Robinson bound~\cite{Vanderstraeten2018,Tran2019,Colmenarez2020}. Moreover,  with the experimental realizations of quantum many-body systems with LR interactions, such as the Rydberg atom arrays~\cite{samajdar2021quantum,yan2022triangular,Semeghini21,Roushan21,yanEmergent2023} and programmable quantum simulators~\cite{verresenPrediction2021,samajdarEmergent2022,yanFully2022,ranFully2022,ranCubic*2023,wangFractionalized2021}, and the magic angle twisted bilayer graphene and 2d moir\'e materials~\cite{tramblyLocalization2010,rafiMoire2011,lopesContinuum2012,tramblyNumerical2012,rozhkovElectronic2016,caoUnconventional2018,caoCorrelated2018,xieSpectroscopic2019,luSuperconductors2019,kerelskyMaximized2019,liaoValence2019,yankowitzTuning2019,yankowitzTuning2019,tomarkenElectronic2019,caoStrange2020,shenCorrelated2020,kevinStrongly2020,soejimaEfficient2020,chatterjeeSkyrmion2022,khalafSoftmodes2020,xieNature2020,khalafCharged2021,pierceUnconventional2021,liaoCorrelated2021,rozenEntropic2021,zondinerCascade2020,saitoIsospin2021,parkFlavour2021,kwanExciton2021,liaoCorrelation2021,kangCascades2021,liuTheories2021,schindlerTrion2022,brillauxAnalytical2022,songMagic2022,linSpin2022,bhowmikBroken2022,huangObservation2022,zhangCorrelated2022,herzogReentrant2022,andreiGraphene2020,stepanovCompeting2021,panThermodynamic2022,zhangMomentum2021,hofmannFermionic2022,panDynamical2021,zhangFermion2022,zhangSuperconductivity2022,zhangQuantum2022,chenRealization2021,linExciton2022,huangIntrinsic2022,huangEvolution2023}, there grows significant motivation to examine the novel properties of such systems.

	\textcolor{black}{In the meantime, the exploration of entanglement properties, predominantly the scaling behavior of entanglement entropies, has emerged as a pivotal subject within the field of condensed matter physics. Such a focus derives from its capacity to effectively characterize various quantum states of matter, as well as the transitions between them, including those beyond the conventional Landau-Ginsburg-Wilson paradigm.
	Preceding investigations of EE have mainly concentrated on short-range systems~\cite{Tagliacozzo2009,Gliozzi_2010,Metlitski2011entanglement,emidioUniversal2022,Kallin_2014}. When considering long-range systems, intriguing results have been found, including the violation of area law scaling in $1d$~\cite{KoffelEntanglement2012,Zehan2021}, the upper bound of area law coefficient~\cite{Gong2017} and so on. The unbiased scaling behavior of EE including sub-leading corrections in $2d$ long-range systems, however, has yet to be explored, primarily due to computational limitations.
	Nevertheless, the advent of non-equilibrium methods for measuring R\'enyi entanglement entropies~\cite{albaOut2017,Emidio2020,zhaoMeasuring2022,zhaoScaling2022,panComputing2023} has opened new avenues for research. These advancements have made the precise detection of such properties in $2d$ long-range systems increasingly feasible.}
	
	In this paper we \textcolor{black}{examine both the entanglement and critical properties for such systems in 2$d$} by studying
	the LR spin-1/2 antiferromagnetic bilayer Heisenberg model with intra-layer power-law decaying ($\frac{1}{r_{ij}^{\alpha}}$) interactions. For the SR case with only nearest-neighbor couplings, the system  has a $SU(2)$ symmetry and by tunning the ratio of inter- and intra-layer couplings the system undergoes a $(2+1)d$ $O(3)$ continuous phase transition from N\'eel state to a Dimer product state~\cite{bilayer2015,wangScaling2022}. With intra-layer LR interactions, the systems still display the N\'eel-to-Dimer phase transition, however the transition is now modified. Here we aim to explore  the critical and entanglement properties of this model systematically via field theory analysis, large-scale unbiased quantum Monte Carlo (QMC) simulations and the nonequilibrium incremental algorithm for measuring EE~\cite{Emidio2020,zhaoScaling2022,zhaoMeasuring2022,demidioUniversal2022,panComputing2023,liaoTeaching2023,liaoControllable2023}. 
	
	Our findings reveal the critical exponents at the QCPs of the $2d$ LR bilayer Heisenberg model vary with $\alpha$ and can be classified into three regimes: the LR regime ($\alpha<\frac{10}{3}$) with a Gaussian fixed point, the SR regime ($\alpha>\alpha_{c}$ with $\alpha_c=3.9621$) with WF critical exponents, and a LR non-Gaussian regime ($\frac{10}{3}<\alpha<\alpha_c$) where critical exponents vary continuously from the LR to SR regimes. In addition, we find that both the area law and the corner correction coefficients of EE at the critical points decrease as LR interactions become stronger. However, in the N\'eel state, only the area law coefficient decays as $\alpha$ becomes smaller, and the logarithmic corrections increases as $\alpha$ decreases, attributing to the variation of anomalous dynamical exponent $z(\alpha)$ as a function of $\alpha$~\cite{DiesselGeneralized2022,songDynamical2023}.

	\begin{figure}[htp!]
		\includegraphics[width=0.99\columnwidth]{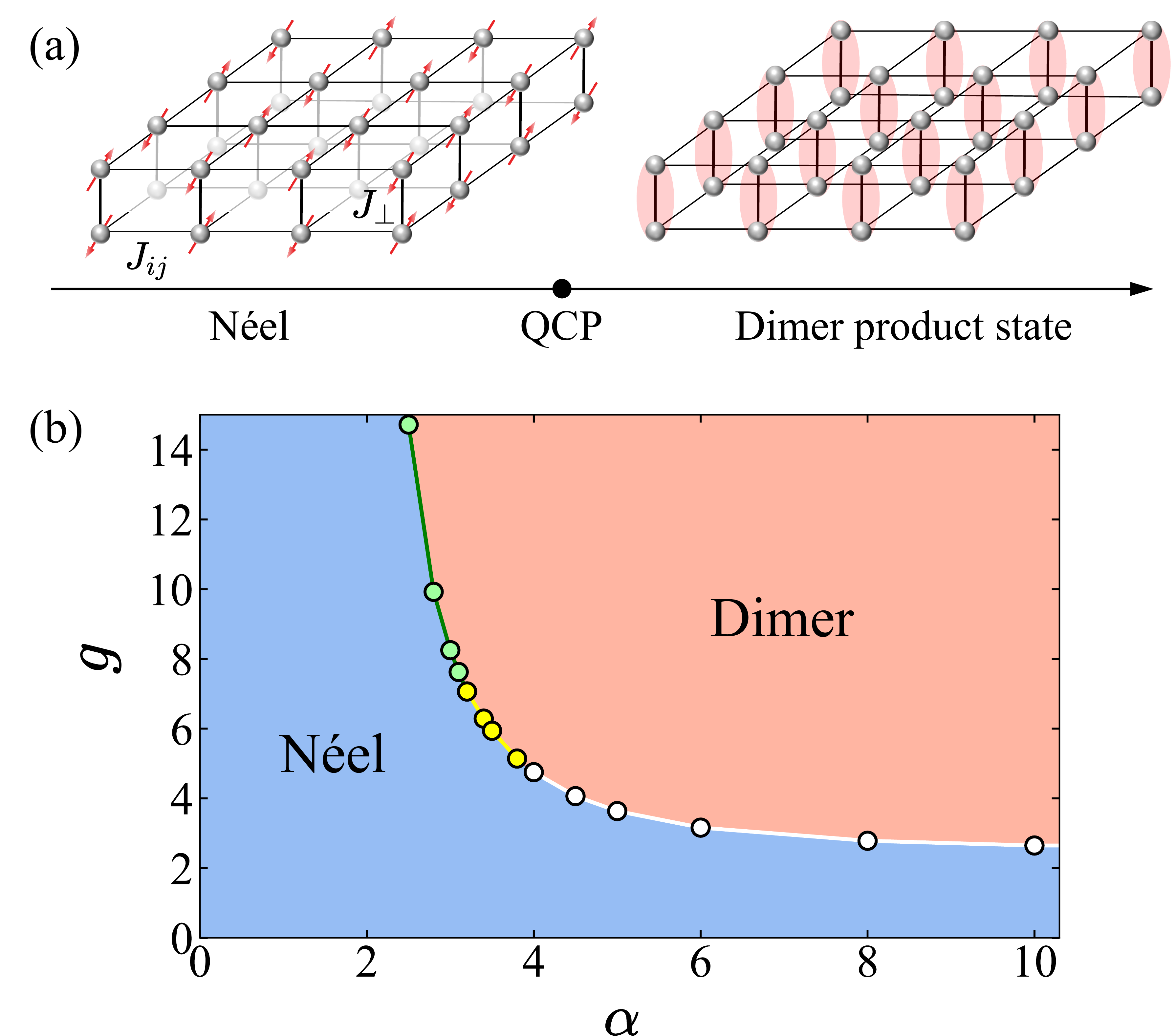}
		\caption{\textbf{Model and Phase diagram of the $2d$ LR antiferromagnetic Heisenberg bilayer}. (a) The bilayer antiferromagnetic model with interlayer interaction $J_{\perp}$ and intralayer LR interaction $J_{ij}$. (b) The ground state phase diagram of model expanded by the axes of $\alpha$ and $g=\frac{J_{\perp}}{J}$, obtained from QMC simulation and finite size analysis, as exemplified in Figs.~\ref{fig:fig3}(a), (b) and in SM~\cite{suppl}. The N\'eel phase spontaneously breaks the spin $SU(2)$ symmetry and the Dimer phase is a product state without symmetry breaking. The QCPs separating them belong to the $(2+1)d$ $O(3)$ universality when $\alpha>\alpha_c=3.9621$, a non-Gaussian fixed point when $\frac{10}{3}<\alpha<\alpha_c$ and a Gaussian fixed point when $\alpha<\frac{10}{3}$. QCPs in LR Gaussian, LR non-Gaussian and SR regions are denoted by green, yellow and white dots (where the simulations are performed) and lines, respectively.}
		
		\label{fig:fig1}
	\end{figure}
	
\begin{figure}[htp!]
		\includegraphics[width=0.95\columnwidth]{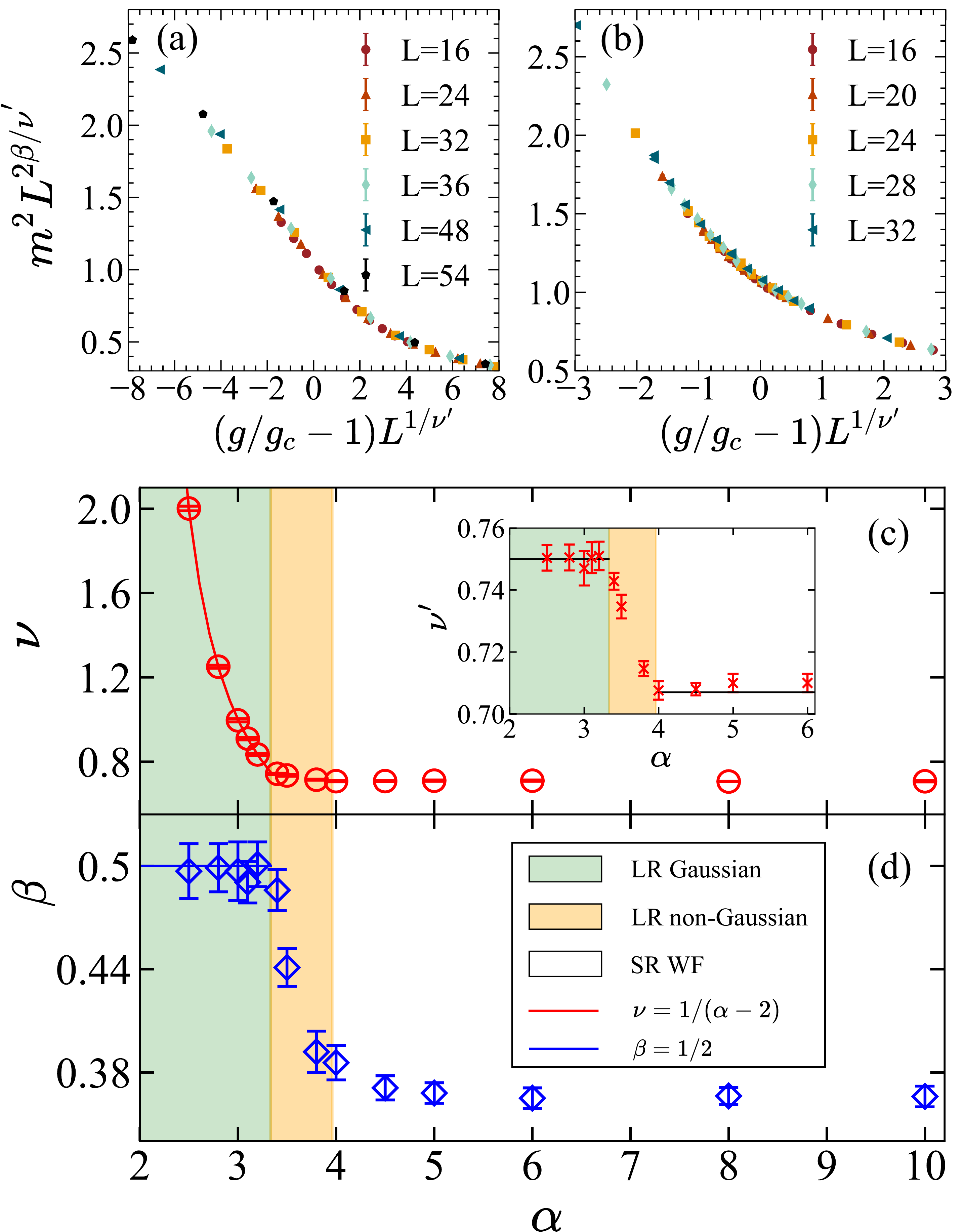}
		\caption{\textbf{Data collapse and the critical exponents $\nu$ and $\beta$ obtained in $\alpha\in(2,10]$.} (a) Data collapse at $\alpha=8$ (SR case) with critical exponents $\nu=0.706(1)$ and $\beta=0.366(5)$. (b) Data collapse at $\alpha=3$ (LR case with Gaussian fixed point above the upper critical dimension) with critical exponents $\nu^{\prime}=0.747(4)$, $\beta=0.497(7)$ and $\nu=\nu^{\prime}\frac{d}{d_{\text{uc}}}$=$0.996(7)$.  The green shaded area in (c) and (d) denote the LR Gaussian regime ($\alpha<\frac{10}{3}$) where $d=2$ is larger than the upper critical dimension $d_{\text{uc}}$. In the region of $\frac{10}{3}<\alpha<\alpha_c$ (yellow shaded area), the system is in a non-Gaussian fixed point and when $\alpha>\alpha_c$ (white area) the critical exponents become their SR $(2+1)d$ $O(3)$ WF values~\cite{wangScaling2022}.}
		\label{fig:fig3}
	\end{figure}

	\noindent{\textcolor{blue}{\it Model and method.}---}
	We consider the spin-1/2 Heisenberg model on a square-lattice bilayer with antiferromagnetic LR intra-layer coupling $J_{ij}$ and inter-layer coupling $J_{\perp}$ {\textcolor{black} {with periodic boundary conditions,}} as shown in Fig.~\ref{fig:fig1}(a). \textcolor{black}{The Hamiltonian is}
	\begin{equation}
		H =\sum_{i\neq j}J_{ij} (\vec{S_{i,1}}\cdot	\vec{S_{j,1}}
		+\vec{S_{i,2}}\cdot \vec{S_{j,2})}
		+ J_{\perp}\sum_i \vec{S_{i,1}}\cdot	\vec{S_{i,2}} ,
   \label{eq:eq1}
	\end{equation}
	where
	$
	J_{ij}=J\frac{(-1)^{|x_i+y_i-x_j-y_j+1|}}{|\vec{r_{i}-\vec{r_{j}}|^{\alpha}}}
	$
	is defined as  a staggered coupling parameter which introduces no frustrations and subscripts $1,2$ denote different layers. We denote $g=J_{\perp}/J$ as the tuning parameter and previous studies have shown that when there are only nearest-neighbor intra-layer interactions, $g_{c}=2.5220(1)$ separates N\'eel ordered phase with the Dimer product phase and this transition is in the $(2+1)d$ $O(3)$ universality class~\cite{bilayer2015,wangScaling2022}. \textcolor{black}{We consider the Ewald-modified coupling parameter~\cite{FukuiOrder2009,flores2015finite,KoziolQuantum2021,zhao2023finitetemperature} in the form of
	$
	\tilde{J}_{ij}=\sum_{m,n=-\infty}^{\infty}\frac{(-1)^{|x_i+y_i-x_j-y_j+1|}}{|\vec{r_i}-\vec{r_j}+mL_{x}\vec{e_{x}}+nL_{y}\vec{e_{y}}|^{\alpha}}
	\label{eq:eq2}
	$
	to tackle the strong finite-size effect in the LR Gaussian regions.} In practice we truncate the summation at $m,n=\pm 500$ which is sufficient to obtain a good finite-size scaling (FSS) behavior as exemplified in Fig.~\ref{fig:fig3}(a) and (b). At larger $\alpha$, we find the original coupling $J_{ij}$ can also obtain converged results.
	
	We also notice that the FSS forms should be modified when the system enters the LR Gaussian region where the system's spatial dimension $d$ is greater than the upper critical dimension $d_{\text{uc}}$. In our case, as will be discussed in the field theory analysis section, $d_{\text{uc}}=\frac{3}{2}(\alpha-d)$ and the system thus enters the LR Gaussian regime when $\alpha<\frac{10}{3}$. Therefore, we write the scaling function that unifies both cases in data collapse to extract the critical exponents $\nu$ and $\beta$ as~\cite{BERCHE2012,Kenna_2013,Kenna_2014,flores2015finite,KoziolQuantum2021,Eduardo2021,zhao2023finitetemperature,Langheld2022,Bertrand2022}	
	\begin{equation}
		m^2 \sim L^{-2 \beta/\nu'} \cdot f\left[L^{1/\nu'}\left(g-g_c\right)\right], \quad g \sim g_c
		\label{data-collapse},
	\end{equation}
	with 
	$
	\nu^{\prime}=\left\{\begin{array}{l}
		\nu \text { for } d<d_{\mathrm{uc}} \\
		\nu \frac{d_{\mathrm{uc}}}{d} \text { for } d>d_{\mathrm{uc}}.
	\end{array}\right.
	$
	Here $\langle m^2 \rangle$ is the square of the N\'eel order parameter $m=\frac{1}{N}\sum_{\vec{r}}(-1)^{r_x+r_y}S^{z}_{\vec{r}}$ and $\nu$ is the actual correlation length exponent. \textcolor{black}{The phase transition point $g_c$ can be located by crossing points of Binder ratios $U(g,L) = \frac{5}{2}(1-\frac{1}{3} \frac{\langle m^4 \rangle}{\langle m^2 \rangle^{2}})$ for various system sizes. The scaling of the Binder ratio crossing points follow the relation of $g^{*}(L)=aL^{-b}+g_c$. In practice, we set $g_c$, $\nu^{\prime}$ and $\beta$ as free parameters and adapt a stochastic data collapse scheme to determine their values accurately~\cite{suppl,yanFully2022,wang2023emus}.} Representative results are shown in Fig.~\ref{fig:fig3}(a) and (b). We have simulated linear system sizes up to $L=54$ in SR regime and up to $L=32$ in LR regime, with the inverse temperature $\beta=L$.
	
	\noindent{\textcolor{blue}{\it Field theory analysis.}---}
	To analyze the QCPs, we consider the following quantum field theory 
	\begin{equation}\label{action}
		S=S_{G}+S_{\text{int}}
	\end{equation}
	with the Gaussian term 
$		S_{G}=\int d\tau dx^d (\sum_i \partial_{\tau}\phi^i(x,\tau))^2+\int d\tau dx^d dx'^{d}\frac{\sum_i \phi^i(x,\tau)\phi^i(x',\tau)}{|x-x'|^{d+\sigma}}$
	and the interaction term
$		S_{int}=\lambda \int d\tau dx^d (\sum_i(\phi^i(x,\tau))^2)^2$.
	\textcolor{black}{It is well know that the $(2+1)d$ short range bilayer anti-ferromagnetic model is in the classical 3D Heisenberg universality class~\cite{SandvikOrder1994}. By universality, the model and the field theory written in Eq.~\eqref{action} (with the second term in Eq.~\eqref{action} replaced by the corresponding local term $\int dx^d dt (\nabla \phi^i(x))^2$) share the same critical behaviour. In particular, the scalar field $(\phi^1,\phi^2,\phi^3)$ can be identified as the expectation value of three component vector $\langle \vec{S} \rangle$. To describe the long-range model, one should instead include the non-local spatial interaction term in Eq.~\eqref{eq:eq1}.}  
	To match the lattice model, we set $\alpha=d+\sigma$. Under the scaling transformation
$		\tau\rightarrow s^{z} \tau, \quad x \rightarrow s x, \quad  \phi^i \rightarrow s^{-\Delta_{\phi}} \phi^i$,
the Gaussian action $S_{G}$ remains invariant if we choose
$		z^G=\frac{\alpha-d}{2},\quad \Delta^G_{\phi}=\frac{3d-\alpha}{4}$.
	After fixing $z$ and $\Delta_{\phi}$, the coupling constant transforms as 
	\begin{equation}
		\lambda \rightarrow s^{\frac{1}{2} (3 \alpha -5d)} \lambda.
	\end{equation}
	When $\alpha<\frac{5d}{3}$, the $\phi^4$ term is irrelevant, the Gaussian fixed point is stable. When $\alpha>\frac{5d}{3}$, the $\phi^4$ term is relevant, which triggers RG flow towards an IR fixed point. For a fixed $\sigma$, we define the upper critical dimension to be
$	 d_{uc}=\frac{3}{2}\sigma$.
	At this IR fixed point, the dynamical critical exponents $z^{\text {IR}}(\alpha)$ and the scaling dimension $\Delta^{\text{IR}}_{\phi}(\alpha)$ will be renormalized. They clearly depends on $\alpha$, even though the precise form of the dependence are not known. This in principle can be studied using quantum field theory techniques by treating $\epsilon=\alpha-\frac{5d}{3}$ as the perturbation parameter (a similar study of the finite temperature LR model was famously done in Ref.~\cite{FisherCritical1972}).
	We expect at some $\alpha_c$, the critical point of the LR models becomes equivalent to the $(2+1)d$ SR Heisenberg model. This will be confirmed by our numerical study later. The crossover from LR to SR happens when $z^{\text {IR}}(\alpha)=1$. 
	
	We now focus on the $d=2$ case. To calculate $\alpha_c$, we need to consider perturbation around the SR model. That is, we consider the following action~\cite{acknow-slava}
	\begin{equation}
		S=S_{\text{CFT}}[\hat{\phi}^i(x)]+\int d\tau dx^2 dx'^{2}\frac{\sum_i\hat{\phi}^i(x,\tau)\hat{\phi}^i(x',\tau)}{|x-x'|^{\alpha}}.
	\end{equation}
	Here $S_{\text{CFT}}$ formally denotes the action of the SR model at criticality, which corresponds to a conformal field theory (CFT). \textcolor{black}{This CFT has been well studied. In particular, the CFT action has a scaling symmetry with $(t,x)\rightarrow  (s t, s x)$. Also, under this system, the scalar field transforms as $s^{-\Delta_{\phi}} \hat{\phi}(s t, t x)$, with scaling dimension $\Delta_{\phi}\approx 0.51892$ \cite{Chester:2020iyt}. The LR term is relevant when $\alpha<\alpha_c=3.9621$.}  The $\alpha=4$ model is another special point. When $\alpha=4$, the action Eq.~\eqref{action} is precisely the SR action after Fourier transformation. The point is therefore in the SR region, which is consistent with $\alpha_c=3.9621$.
	
	\noindent{\textcolor{blue}{\it Critical behavior.}---}
	We first use the crossing points of Binder ratios $g^{*}(L)$ to locate $g_c$ by fitting to the relation $g^{*}(L)=aL^{-b}+g_{c}$. Then we set $g_c$ as a free value around the fitted value and perform  data collapse to determine the values of $g_c$, $\nu'$ and $\beta$ according to Eq.~\eqref{data-collapse}, and the results are exemplified in Fig.~\ref{fig:fig3}(a) and (b) for the cases of $\alpha=3$ and 8. We employ the stochastic data collapse approach to obtain high accuracy exponents, the detailed description and examples are given in the Supplemental Materials (SM)~\cite{suppl}. The value of $\nu$ can then be calculated from the relation $\nu'=\frac{d_{\text{uc}}}{d}\nu$. Note that the expected value of $\nu$ in the LR Gaussian regime is $\nu=\frac{1}{\alpha-d}$ and $d_{\text{uc}}=\frac{3}{2}(\alpha-d)$, so $\nu'$ will take the value of 0.75 in the entire Gaussian regime.
	
	We then plot the obtained critical exponents $\nu$ and $\beta$ versus the interaction exponent $\alpha$. As shown in Fig.~\ref{fig:fig3} (c) and (d), when $\alpha>\alpha_c$ the critical exponents take  the SR values of the $(2+1)d$ $O(3)$ universality class with $\nu=0.706(1)$ and $\beta=0.366(5)$~\cite{bilayer2015,wangScaling2022}.  When $\alpha<\frac{10}{3}$ the system enters the LR Gaussian regime with $\nu=\frac{1}{\alpha-2}$ and $\beta=0.5$. Between these two regimes, the critical exponents $\nu$ and $\beta$ varies continuously with $\alpha$ from the SR to the LR values.  {\textcolor{black} The deviation of $\beta$ from its SR value at $\alpha=4$ is due to strong finite-size effects near the crossover at $\alpha=\alpha_c$.}

	\noindent{\textcolor{blue}{\it Entanglement entropy.}---}
	The violation of area law scaling of EE in $1d$ quantum spin chains with LR interactions  has been observed via density matrix renormalization group (DMRG)~\cite{KoffelEntanglement2012,Zehan2021}. {\textcolor{black}However, apart from few works~\cite{Gong2017,kuwahara2020area} discussing the reliability of area law scaling of EE in LR systems, the important scaling form of EE for $2d$ LR systems is unexplored.} Meanwhile, with the fast development of QMC  algorithms for EE computation~\cite{Emidio2020,zhaoScaling2022,zhaoMeasuring2022,demidioUniversal2022,panComputing2023,liaoTeaching2023,song2023resummationbased}, the R\'enyi EE now can be measured with high precision and efficiency both in phases and at the critical points in $2d$ systems.
	\begin{figure}[htp!]
		\includegraphics[width=0.99\columnwidth]{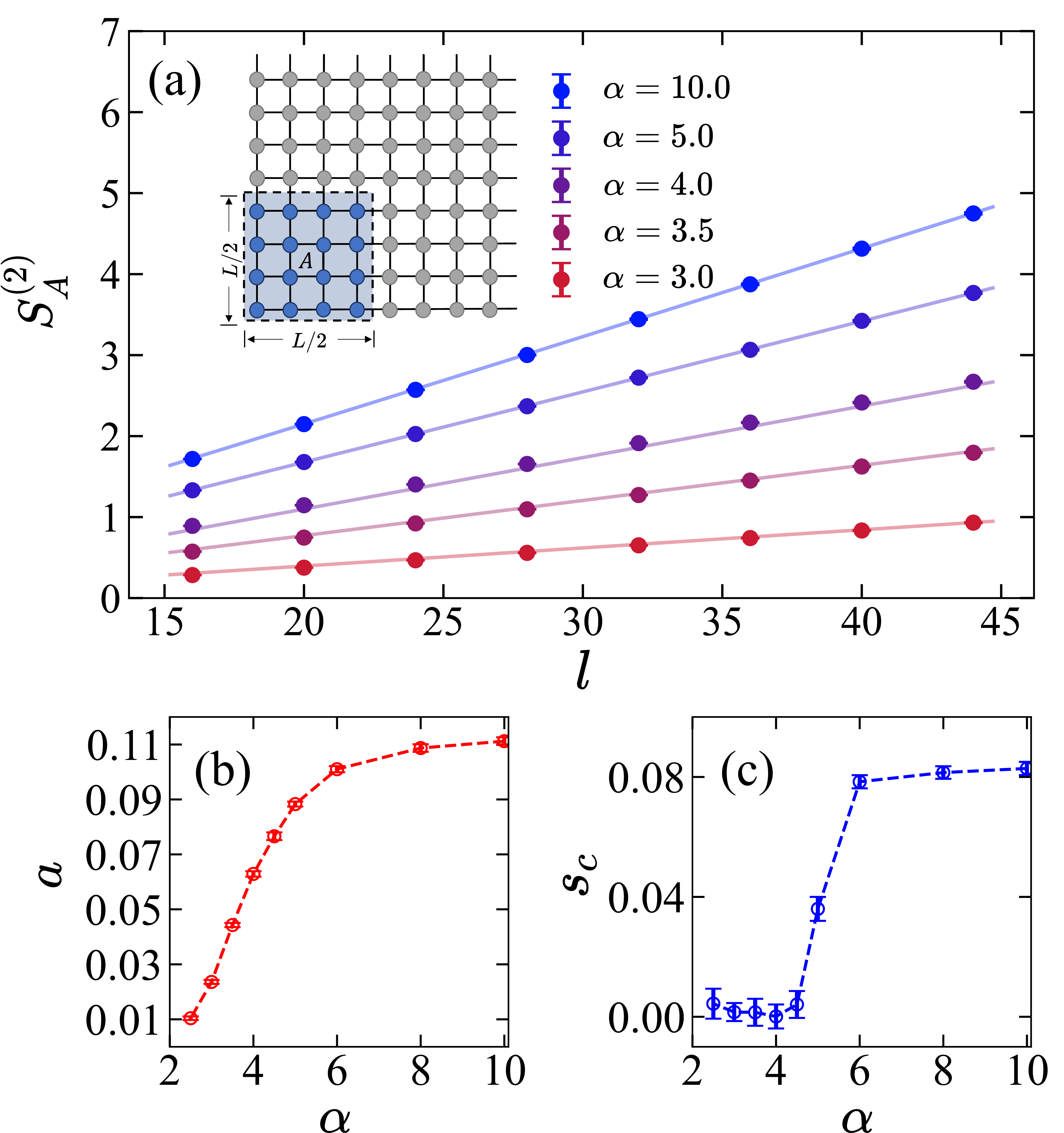}	
		\caption{\textbf{EE at the QCPs.} $S^{(2)}_A$ and its scaling behavior at the QCPs shown in Fig.~\ref{fig:fig1} (b) \textcolor{black}{with system sizes $L=8,10,12,...,22$} (a) R\'enyi entropies versus the system sizes for different interaction power $\alpha$. The inset shows the square entanglement region $A$ with boundary length $l=2L$. (b) The area law coefficient $a$ versus $\alpha$. (c) The log-coefficient $s_c$ versus $\alpha$.}
		\label{fig:fig4}
	\end{figure}
	For $2d$ SR systems, EE takes the form of
	\begin{equation}
		\label{scaling-ee}
		S(l)=al-s_c\ln l +c,
	\end{equation}
	where $l$ is the length of the entanglement boundary and the logarithmic term arises from the contribution of sharp corners on the boundary. For SR models, the value of $s_c$ is universal and $s_c\approx 0.081$ for four $\frac{\pi}{2}$ corners at the $(2+1)d$ $O(3)$ criticality~\cite{zhaoScaling2022,zhaoMeasuring2022,Kallin_2014}. {\textcolor{black} However, for LR systems the scaling forms of EE have not been explored, especially at their QCPs.} We measure the 2nd R\'enyi EE $S^{(2)}_A$ at the QCPs in Fig.~\ref{fig:fig1}(b). We choose a $\frac{L}{2}\times \frac{L}{2}$ square region as $A$ as shown in the inset of Fig.~\ref{fig:fig4}(a) and use the scaling relation defined in Eq.~\eqref{scaling-ee} to fit our results. The fitting results are shown in Fig.~\ref{fig:fig4}(b) and (c), and one finds as $\alpha$ decreases the area law coefficient gradually decays to a small value, which indicates that quantum entanglement at the criticality becomes weaker when the interactions becomes more long-ranged. This can be understood by the fact that the critical points are described by a LR Gaussian theory for $\alpha<\frac{10}{3}$ and \textcolor{black}{the system becomes more mean-field like when LR interactions are enhanced}. The corner corrections drop rapidly to zero as the system goes into the LR region $\alpha<\alpha_c$, which can be understood that strong LR interactions trivialize the geometry of sharp corners and make the EE less sensitive to the shape of the entanglement boundary.
	\begin{figure}[htp!]
		\includegraphics[width=0.99\columnwidth]{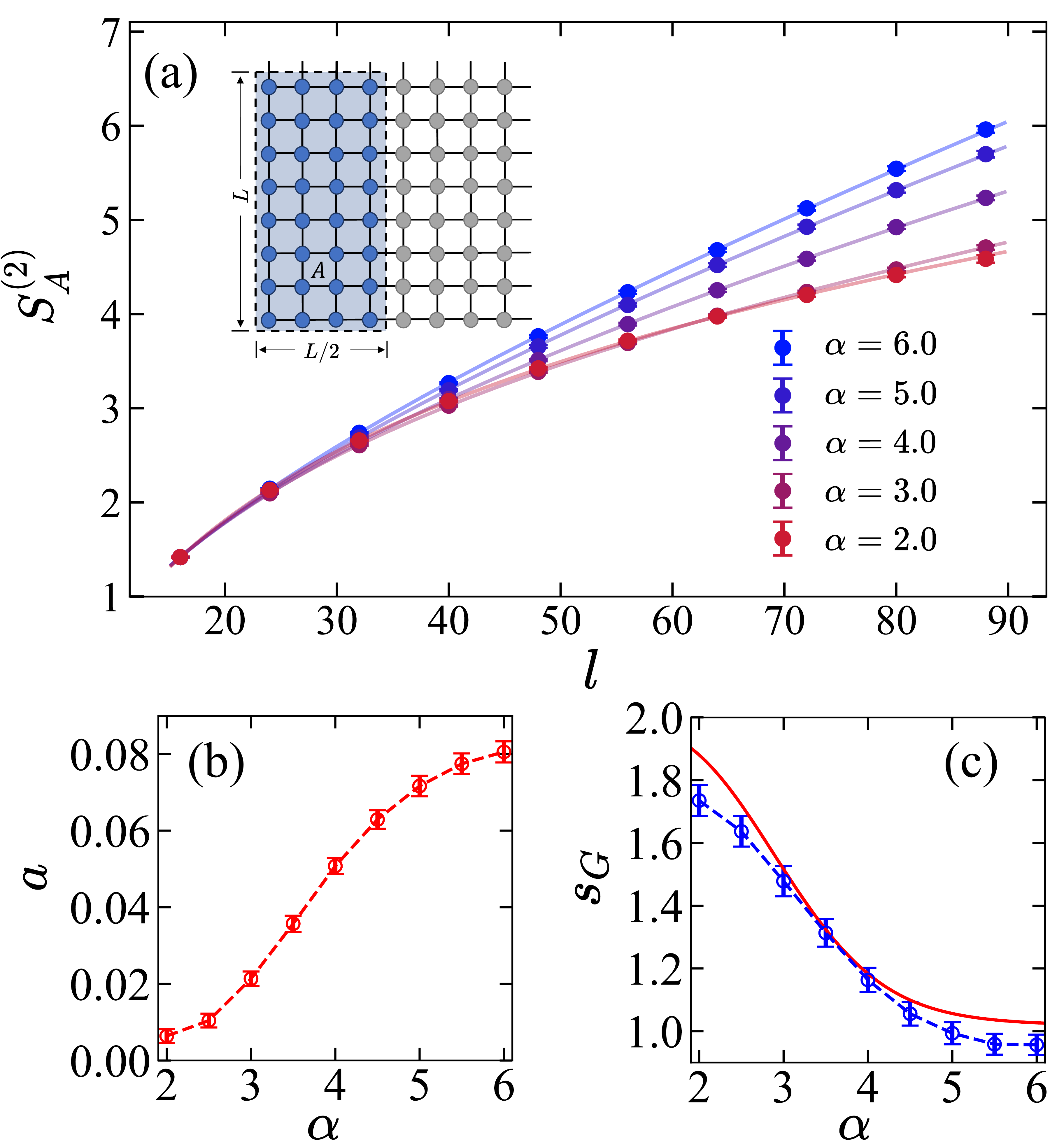}
		\caption{\textbf{EE inside the N\'eel phase.} $S^{(2)}_A$ and its scaling behavior for single-layer LR Heisenberg model i.e. $g=0$ in Eq.~\eqref{eq:eq1} in the N\'eel phase \textcolor{black}{with system sizes $L=8,12,16,...,44$} (a) R\'enyi entropies versus the system sizes for different interaction power $\alpha$. The inset shows the cylinder entanglement region $A$ with boundary length $l=2L$. For clarity, $S^{(2)}_{A}$  is modified with a constant so that  $S^{(2)}_{A}(l=16)$ is the same for every $\alpha$. (b) The area law coefficient $a$ versus $\alpha$. (c) The logarithmic coefficient $s_G$ versus $\alpha$. The red line shows the result $s_G=\frac{N_{G}(d-z(\alpha))}{2}$ with $z(\alpha)$ obtained from spin wave theory and QMC data ($L=64$) in Ref.~\cite{songDynamical2023}.}
		\label{fig:fig5}
	\end{figure}	
	
	We also examine the scaling of EE in the N\'eel phase in Fig.~\ref{fig:fig5}. As shown in the inset of Fig.~\ref{fig:fig5}(a), we choose region $A$ to be the $L\times \frac{L}{2}$ partition with smooth boundary. In this setting, the logarithmic corrections are purely from the interplay between gapless Goldstone modes and restoration of the symmetry in a finite system~\cite{Irenee2017,Metlitski2011entanglement}. With the addition of power-law decaying LR interactions, the dynamic exponent $z(\alpha)$ of the N\'eel state is modified~\cite{songDynamical2023,DiesselGeneralized2022} as well as the structure of tower of state~\cite{Irenee2017,Metlitski2011entanglement}. In this case, the $s_c$ in Eq.~\eqref{scaling-ee} needs to be replace by $-s_G=-\frac{N_{G}(d-z(\alpha))}{2}$ and $N_G$ is the number of Goldstone modes.
	We thus fit our results of EE with Eq.~\eqref{scaling-ee} and the fitting results are shown in Fig.~\ref{fig:fig5}(b) and (c). We find the area law coefficient $a$ also decays as the LR interactions get stronger, whereas the logarithmic coefficient $s_G$ increases from the SR value $s_G=\frac{N_G}{2}=1$ to $s_G\approx 1.76$ as $\alpha$ decreases from 6 to 2. We substitute the data of $z(\alpha)$ from previous QMC and spin wave analysis~\cite{songDynamical2023} into $s_G=\frac{N_{G}(d-z(\alpha))}{2}$ and find good agreement with the fitted $s_G$, as shown in Fig.~\ref{fig:fig5}(c). The derivation in the SR and LR Gaussian regimes attribute to finite-size effects. 
	
	\noindent{\textcolor{blue}{\it Discussions.}---}
	In this work, {\textcolor{black}we address the important open questions regarding the critical exponents and scaling of EE in $2d$ LR quantum many-body system.} Through large-scale QMC simulations, we obtained precise critical exponents in the LR Gaussian regime with a Gaussian fixed point, the SR regime with WF exponents, and an LR non-Gaussian regime where the critical exponents vary continuously from LR to SR values. Our investigation of R\'enyi EE has revealed highly nontrivial features both along the QCP line and in the N\'eel phase, {\textcolor{black}in that, as the interaction becomes longer-ranged, the area law contribution in EE gradually vanishes, while the log-correction is enhanced in the N\'eel phase due to the anomalous dependence of the dynamical exponent $z(\alpha)$ on $\alpha$.} Our results have important implications for future theoretical and experimental investigations of LR interacting quantum many-body systems, including Rydberg atom arrays~\cite{samajdar2021quantum,yan2022triangular,Semeghini21,Roushan21,yan2022triangular,yanEmergent2023}, programmable quantum simulators~\cite{verresenPrediction2021,samajdarEmergent2022,yanFully2022,ranFully2022,wangFractionalized2021}, and magic angle twisted bilayer graphene and $2d$ moir\'e materials. 
	
	\vspace{0.2cm}
		{\it{Acknowledgment.-}} We thank Subir Sachdev, Fabien Alet, Fakher Assaad, Kai Sun, Michael Scherer, Slava Rychkov, Lukas Janssen, Meng Cheng and Youjin Deng for valuable discussions on the related topic. MHS, JRZ and ZYM acknowledge support from the Research Grants Council (RGC) of Hong Kong Special Administrative Region (SAR) of China (Projects Nos. 17301420, 17301721, AoE/P-701/20, 17309822 and HKU C7037-22G), the ANR/RGC Joint Research Scheme sponsored by the RGC of Hong Kong SAR of China and French National Research Agency (Project No. A\_HKU703/22) and the HKU Seed Funding for Strategic Interdisciplinary Research “Many-body paradigm in quantum moiré material research”. The work of JR are supported by the Huawei Young Talent program in Institut des Hautes \'{E}tudes Scientifiques. The authors also acknowledge the Tianhe-II platform at the National Supercomputer Center in Guangzhou, the HPC2021 system under the Information Technology Services
	and the Blackbody HPC system at the Department of Physics, University of Hong Kong for their technical support and generous allocation of CPU time.
	
	\bibliography{bibtex}
	
%%%%%%%%%%%%%%%%%%%%%%%%%%%%%%%%%%%%%%%%%%%%%%%%%%%%%%%%%%%%%%%%%%%%%%%%%%
% SUPPLEMENTARY MATERIALS
%%%%%%%%%%%%%%%%%%%%%%%%%%%%%%%%%%%%%%%%%%%%%%%%%%%%%%%%%%%%%%%%%%%%%%%%%%
\clearpage
\onecolumngrid

%\appendix
\setcounter{equation}{0}
\setcounter{figure}{0}
\setcounter{table}{0}
\setcounter{page}{1}
\makeatletter
\renewcommand{\theequation}{S\arabic{equation}}
\renewcommand{\thefigure}{S\arabic{figure}}
\setcounter{secnumdepth}{3}	

\begin{center}
	\bf \uppercase{Supplementary Materials for ``Quantum criticality and entanglement for two dimensional long-range Heisenberg bilayer''}
\end{center}
\vspace{2\baselineskip}

\twocolumngrid

\section{QMC implementation}
\subsection{Stochastic data collapse}
We develop a stochastic data collapse procedure similar to Ref.~\cite{yanFully2022,wang2023emus} to accurately estimate the critical points and exponents. A polynomial curve is fitted from the data points to be collapsed of different system sizes $L$. The goodness of fit is associated with the quality of data collapse and thus the accuracy of estimated critical exponents. We use the R square value $R^2$ to evaluate the variation between our actual data and fitting curve. The definition of $R^2$ is $R^2 = 1-\frac{S_{res}}{S_{tot}}=1-\delta$,
with $S_{res}=\sum_{i=1}^{n}w_{i}\left(y_{i}-\hat{y}_{i}\right)^{2},S_{tot}=\sum_{i=1}^{n}w_{i}\left(y_{i}-\bar{y}\right)^{2}$. $\delta$ can be regarded as the error of fit and is to be minimized. $\hat{y}_i$ is the $y$-value of the scaled data point, e.g. $m^2L^{2\beta/\nu^{\prime}}$ and $y_i$ is that of the fitting curve at the same $x$-value. $\bar{y}$ is the mean value of all points $y_i$ of the fitting curve. $w_i$ is the weight which has a larger value near the critical points which emphasize a high collapse quality in the critical region. In general, $S_{res}$ measures the deviation between the actual data and the fitting curve, and $S_{tot}$ measures variance of the fitting curve itself.
\begin{figure}[htp!]
	\includegraphics[width=0.8\columnwidth]{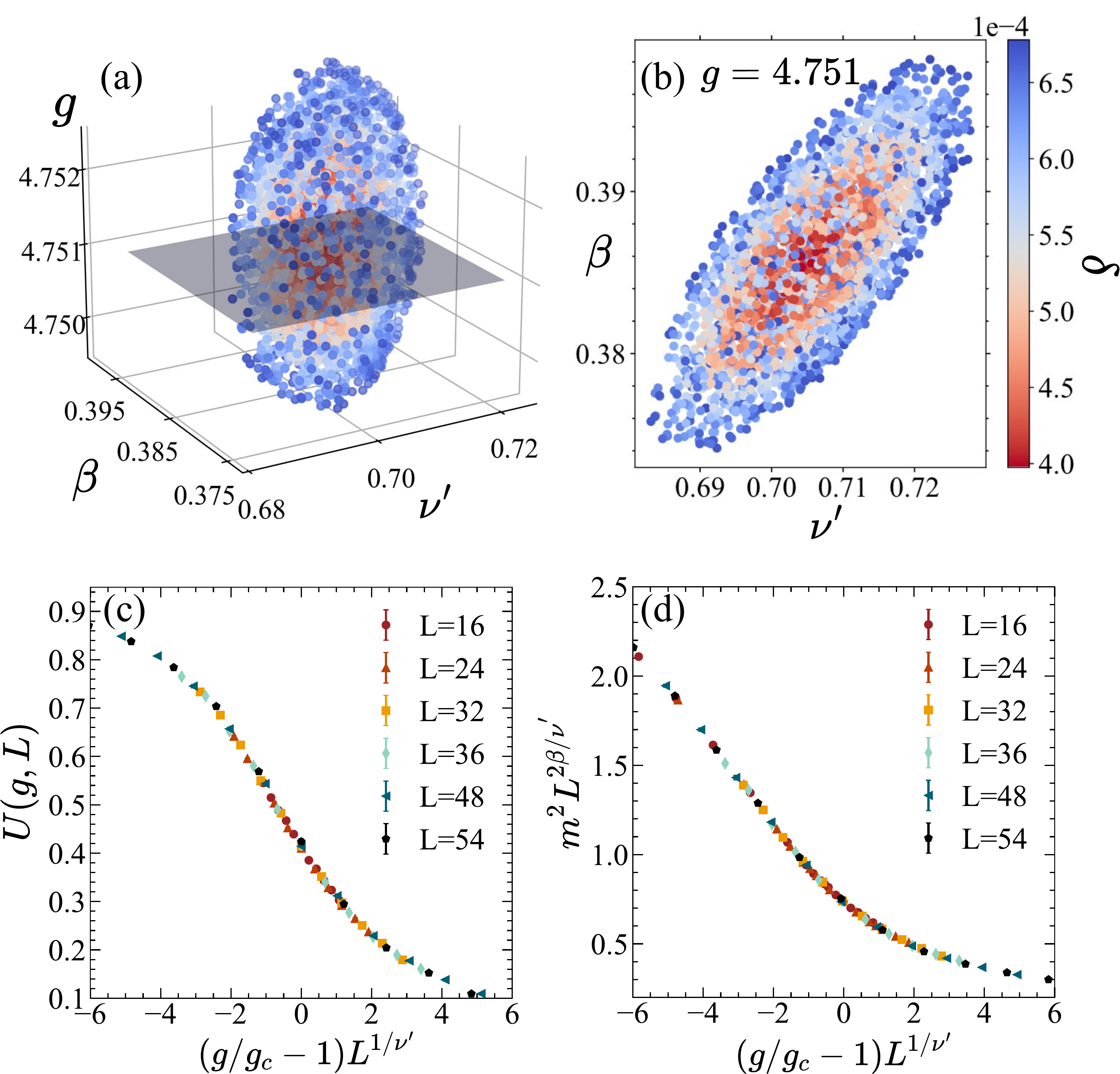}
	\caption{\textbf{Example of stochastic data collapse at $\mathbf{\alpha=4.0}$}. (a) Distribution of fitting error $\delta$ in the space spanned by $g$, $\beta$ and $\nu^{\prime}$ axes. (b) The top view of the shaded tangent plane $g=4.751$ in (a). (c) and (d) show the collapses of Binder ratio and square of staggered magnetization with $g_c=4.751$, $\beta=0.385$ and $\nu^{\prime}=0.707$ obtained by minimizing fitting error $\delta$.}
	\label{fig:sm1}
\end{figure}
One can set $g$, $\beta$ and $\nu^{\prime}$ as free parameters and the stochastic process is done in the three dimensional space spanned by the three parameters. To begin, a random set of parameter is proposed and fitted by a polynomial curve. Fitting error $\delta$ is calculated. Then one stochastic ally move three parameters in the 3-d space as shown in Fig.~\ref{fig:sm1}(a) while recording the fitting error $\delta$. After enough steps, the parameter set with the smallest error is chosen to be the best estimation. The distribution of $\delta$ is shown in Fig.~\ref{fig:sm1}(a) and a top view at plane $g=4.571$ is shown in Fig.~\ref{fig:sm1}(b). The collapses of Binder ratio and square of staggered magnetization using $g_c$, $\nu^{\prime}$ and $\beta$ with the smallest fitting error $\delta$ are shown in Fig.~\ref{fig:sm1}(c) and (d) respectively.

\end{document}